\documentclass{SciPost}

\renewcommand{\arraystretch}{1.3}
\hypersetup{
    colorlinks,
    linkcolor={red!50!black},
    citecolor={blue!50!black},
    urlcolor={blue!80!black}
}

\usepackage[bitstream-charter]{mathdesign}
\usepackage[]{placeins}
\DeclareSymbolFont{usualmathcal}{OMS}{cmsy}{m}{n}
\DeclareSymbolFontAlphabet{\mathcal}{usualmathcal}

\fancypagestyle{SPstyle}{
\fancyhf{}
\lhead{\colorbox{scipostblue}{\bf \color{white}Prepared for submission to SciPost Physics Community Reports }}
\rhead{{\bf \color{scipostdeepblue}}}

\fancyfoot[C]{\textbf{\thepage}}
}

\begin{document}
\vspace*{-1cm}
\begin{flushright} 
{\color{magenta}LHCHWG-2025-013}
\end{flushright}
\vspace*{0.5cm}

\pagestyle{SPstyle}

\begin{center}{\Large \textbf{\color{scipostdeepblue}{
    Impact of Higgs-Boson measurements on SMEFT Fits
}}}\end{center}

\begin{center}\textbf{
J. de Blas\textsuperscript{1$\star$},
A. Goncalves\textsuperscript{2$\dagger$},
V. Miralles\textsuperscript{3,4$\circ$},
L. Reina\textsuperscript{2$\S$},
L. Silvestrini\textsuperscript{5$\P$} and
M. Valli\textsuperscript{5$\parallel$}
}\end{center}

\newcommand{\mv}[1]{\textcolor{red}{\bf MV: #1}}

\begin{center}
{\bf 1} Departamento de Física Te\'orica y del Cosmos, Universidad de Granada, Campus de Fuentenueva, E-18071 Granada, Spain\\
{\bf 2} Physics Department, Florida State University, Tallahassee, FL 32306-4350, USA\\
{\bf 3} Departament de F\'isica, Universitat d'Alacant,
Campus de San Vicent del Raspeig, E-03690 \mbox{Alacant}, Spain\\
{\bf 4} School of Physics and Astronomy, University of Manchester, Oxford Road, Manchester M13 9PL, UK\\
{\bf 5} INFN, Sezione di Roma, Piazzale A. Moro 2, I-00185 Rome, Italy
\\[\baselineskip]
$\star$ \href{mailto:deblas@ugr.es}{\small deblas@ugr.es}\,,\quad
$\dagger$ \href{mailto:agoncalvesdossantos@fsu.edu}{\small agoncalvesdossantos@fsu.edu}\,,\quad
$\circ$ \href{mailto:victor.miralles@ua.es}{\small victor.miralles@ua.es}\,,\quad
$\S$ \href{mailto:reina@hep.fsu.edu}{\small reina@hep.fsu.edu}\,,\quad
$\P$ \href{mailto:Luca.Silvestrini@roma1.infn.it}{\small Luca.Silvestrini@roma1.infn.it}\,,\quad
$\parallel$ \href{mailto:Mauro.Valli@roma1.infn.it}{\small Mauro.Valli@roma1.infn.it}
\\[\baselineskip]
\end{center}

\section*{\color{scipostdeepblue}{Abstract}}
\textbf{\boldmath{%
We present current bounds on SMEFT operators that are mainly constrained by Higgs-boson observables, under different assumptions for the flavour structure of the UV theory. We investigate how the accuracy reached through a dedicated Higgs-boson precision physics program is starting to have a major impact in setting a lower bound on the scale of new physics, and we discuss the relevance of considering the scale evolution of the SMEFT coefficients in this context. We compare our results with the literature, pointing out the consistency of the results in spite of the different assumptions adopted in each analysis, and we discuss future steps aimed at improving the accuracy of the fit.
}}

\vspace{\baselineskip}




\vspace{10pt}
\noindent\rule{\textwidth}{1pt}
\tableofcontents
\noindent\rule{\textwidth}{1pt}
\vspace{10pt}


\section{Introduction}
\label{sec:intro}

The Standard Model Effective Field Theory (SMEFT) provides a systematic framework to probe effects of new physics (NP) beyond the Standard Model (SM) under well-defined limited assumptions. In its general form the SMEFT Lagrangian can be written as an extension of the SM Lagrangian by a series of Lorentz-invariant local operators ${\cal O}_i^{(d)}$ of canonical mass dimension larger than four ($d>4$) built of SM fields and respecting the SM gauge symmetry:
\begin{equation}
\label{eq:smeft-lagrangian}
\mathcal{L}_{\mathrm{SMEFT}}=\mathcal{L}_{\mathrm{SM}}+\sum_{d>4}\frac{1}{\Lambda^{d-4}}\sum_{i}C_i^{(d)}\mathcal{O}_i^{(d)},
\end{equation}
where $i$ collectively labels each operator according to its field composition and quantum numbers, $C_i^{(d)}$ denote the corresponding Wilson coefficients, and $\Lambda$ is the cutoff of the effective theory indicatively identified with the ultraviolet (UV) scale of NP. In this work, we truncate the $\mathcal{L}_{\mathrm{SMEFT}}$ to include only dimension-6 operators, consistently retain only their linear order in all calculated observables, and adopt the Warsaw basis introduced in Ref.~\cite{Grzadkowski:2010es}. In the following we will denote the corresponding Wilson coefficients simply by $C_i$ (omitting the $d=6$ index) and will consider their scale dependence ($C_i=C_i(\mu)$) as dictated by the leading-order SMEFT renormalization group evolution from the scale of NP $\Lambda$ to the proper scale of individual observables.

Additional global symmetries may be imposed on the $d>4$ terms of the SMEFT, covering in this way a variety of possible NP models. For this study we consider effective interactions that respect baryon and lepton number, and do not introduce additional sources of CP-violation beyond the SM one. Furthermore, without changing the flavour structure of the SM, one can require NP to satisfy different flavour symmetries which will then be reflected on the $d>4$ SMEFT operators and constrained by a global fit that also includes flavor observables. In this study we will consider the two scenarios in which NP satisfies a $U(3)^5$ or $U(2)^5$ flavour symmetry~\cite{Faroughy:2020ina,deBlas:2025xhe}.

The $C_i$ coefficients can be constrained comparing SMEFT predictions to experimental measurements of various physical observables. In this respect, the SMEFT framework naturally lends itself to investigate correlations among different operators induced by their action on common sets of observables or by renormalization mixing. Once established, correlations can point to specific classes of models. Recent fits of the SMEFT~\cite{Ellis:2020unq,Ethier:2021bye,Garosi:2023yxg,Allwicher:2023shc,Bartocci:2023nvp,Celada:2024mcf,ATLAS:2024lyh,Bartocci:2024fmm,CMS:2025ugn,terHoeve:2025gey,deBlas:2025xhe} have considered a broad set of observables, including electroweak (EW) precision observables, Higgs-boson signal strengths and simplified template cross sections, top-quark rates and asymmetries, Drell-Yan and di-boson rates, as well as flavour-physics observables such as $K$- and $B$-meson mixing and $K$-, $D$-, and $B$-meson decays. 

In this context, it is interesting to focus on the role played by Higgs observables and their constraining power within fits of SMEFT interactions, based on the latest LHC results~\cite{ATLAS:2021vrm,ATLAS:2022vkf,ATLAS:2024lyh,ATLAS:2024fkg,ATLAS:2024ish,CMS:2022dwd,CMS:2025jwz}. In this study we therefore consider those operators that are mainly constrained by Higgs-boson measurements, as determined by recent analyses~\cite{Allwicher:2023shc,Bartocci:2023nvp,Celada:2024mcf,Bartocci:2024fmm,deBlas:2025xhe}, and investigate how the accuracy reached through a dedicated Higgs-boson precision physics program is starting to have a major impact in constraining the corresponding Wilson coefficients and setting a lower bound on the scale of NP. We present individual fits of the corresponding SMEFT coefficients based on the latest experimental results and investigate two different NP flavour-scenarios, namely the case in which the SMEFT effective interactions at the scale $\Lambda$ are $U(3)^5$ or $U(2)^5$ flavour invariant. We will highlight how the two scenarios give origin to different constraints on the $C_i$ coefficients.
Furthermore, in each case, we will highlight the relevance of considering the scale evolution of the $C_i$ coefficients and discuss future steps aimed at controlling it with improved accuracy. Finally, we discuss the consistency of the results presented in recent studies~\cite{Allwicher:2023shc,Bartocci:2023nvp,Celada:2024mcf,Bartocci:2024fmm,deBlas:2025xhe}, taking into account the different assumptions adopted in each analysis.

\section{Fitting framework}
\label{sec:framework}

\begin{table}[ht!] 
\begin{center}
  {\small
  \renewcommand{\arraystretch}{1.5}
  \scalebox{1.}{
  \begin{tabular}{|c|c|c|} 
  \hline
    $\mathcal{O}_{\phi\Box}=(\phi^\dag \phi) \Box(\phi^\dag \phi)$ &
    $\mathcal{O}_{e\phi}^{[pr]}=(\phi^\dag \phi)(\bar l_p \phi e_r )$ & 
    $\mathcal{O}_{e B}^{[pr]}=(\bar l_p \sigma^{\mu\nu} e_r) \phi B_{\mu\nu}$ \\
    $\mathcal{O}_{\phi G}=\phi^\dag \phi\, G^A_{\mu\nu} G^{A\mu\nu}$ & 
    $\mathcal{O}_{u\phi}^{[pr]}=(\phi^\dag \phi)(\bar q_p \widetilde{\phi} u_r )$ & 
    $\mathcal{O}_{u B}^{[pr]}=(\bar q_p \sigma^{\mu\nu} u_r) \widetilde{\phi}\, B_{\mu\nu}$ \\ 
    $\mathcal{O}_{\phi B}=\phi^\dag \phi\, B_{\mu\nu} B^{\mu\nu}$ &
    $\mathcal{O}_{d\phi}^{[pr]}=(\phi^\dag \phi)(\bar q_p \phi d_r )$ & 
    $\mathcal{O}_{e W}^{[pr]}=(\bar l_p \sigma^{\mu\nu} e_r) \tau^I \phi W_{\mu\nu}^I$ \\ 
    $\mathcal{O}_{\phi W}=\phi^\dag \phi\, W^I_{\mu\nu} W^{I\mu\nu}$ & 
    $\mathcal{O}_{\phi ud}^{[pr]}=(\widetilde{\phi}^\dag iD_\mu \phi)(\bar u_p \gamma^\mu d_r)$ & 
    $\mathcal{O}_{uW}^{[pr]}=(\bar q_p \sigma^{\mu\nu} u_r) \tau^I \widetilde{\phi}\, W_{\mu\nu}^I$ \\
    $\mathcal{O}_{\phi WB}= \phi^\dag \tau^I \phi\, W^I_{\mu\nu} B^{\mu\nu}$ &
    & 
    $\mathcal{O}_{uG}^{[pr]}=(\bar q_p \sigma^{\mu\nu} T^A u_r) \widetilde{\phi}\, G_{\mu\nu}^A$ \\
    $\mathcal{O}_G=f^{ABC} G_\mu^{A\nu} G_\nu^{B\rho} G_\rho^{C\mu} $ &  
    $\mathcal{O}_{lequ}^{(1)[prst]}=(\bar l_p^i e_r)\epsilon_{ij}(\bar q_s^j u_t)$ & \\ 
    $\mathcal{O}_{W}=\varepsilon^{IJK} W_\mu^{I\nu} W_\nu^{J\rho} W_\rho^{K\mu}$ & 
    $\mathcal{O}_{lequ}^{(3)[prst]}=(\bar l_p^i \sigma_{\mu\nu} e_r)\epsilon_{ij}(\bar q_s^j \sigma^{\mu\nu} u_t)$ &\\
    \hline
    \end{tabular}
    }
    }
\caption{ Dimension-6 operators of the Warsaw basis~\cite{Grzadkowski:2010es} considered in this study. We denote by $G^A_{\mu\nu}$ ($A=1,\ldots, 8$), $W^I_{\mu\nu}$ ($I=1,2,3$), and $B^{\mu\nu}$ the gauge field strength tensors of the $SU(3)_C$ , $SU(2)_L$, and $U(1)_Y$ SM gauge groups respectively;  by $\phi$ the complex $SU(2)_L$ doublet scalar field of the
    SM; by $q$ and
    $l$ the $SU(2)_L$ left-handed quark and lepton doublets, and by $u,d,e$ the $SU(2)_L$ right-handed quark and lepton singlets, with family indices specified by the superscript in square
    brackets.\label{tab:smeft-operators}}
\end{center}
\end{table}

Among the dimension-6 operators that satisfy the properties outlined in Sec.~\ref{sec:intro}, the ones that are mainly constrained by Higgs-boson observables are listed in Table~\ref{tab:smeft-operators} (see~\cite{Bartocci:2023nvp,Bartocci:2024fmm,deBlas:2025xhe}).\footnote{Ref.~\cite{Allwicher:2023shc} presents a fit within the $U(2)^5$ flavour assumption but does not include Higgs-boson observables. Hence, their results can only be compared with our ones when we omit Higgs-boson observables, see the “noH" column in Table \ref{tab:u2-ind}, showing a good agreement.} They consist of operators containing only scalar and gauge field (often denoted as \textit{bosonic} operators) as well as \textit{dipole} operators, and operators containing one or two fermion currents.

Notice that we do not consider the operator $\mathcal{O}_{\phi}=(\phi^\dagger\phi)^3$ since it does not enter at leading order in the observables considered in this study. In fact, the main effect of this operator is to modify the Higgs-boson self-interactions, and can only be weakly constrained with current Higgs pair-production data~\cite{CMS:2022dwd,ATLAS:2024ish}. Furthermore, although $\mathcal{O}_{\phi}$ can be generated from other bosonic interactions via renormalization, it does not enter in the leading-order SMEFT anomalous dimension matrix, i.e. setting a non-zero value of $C_\phi$ at the UV scale does not contribute to other coefficients at lower scales via one-loop renormalization-group evolution.

We assume that the SMEFT coefficients are generated by NP at the UV scale $\Lambda$, and treat the $C_i(\Lambda)$ coefficients as parameters of the fit. We then derive the Wilson coefficients at a generic scale $\mu_W$ (for convenience chosen to be $\mu_W=M_W$) through leading-order renormalization group (RG) evolution (RGE)~\cite{Jenkins:2013zja,Jenkins:2013wua,Alonso:2013hga}. We work at linear order in the SMEFT dimension-6 coefficients and write the running of the Wilson coefficients as:
\begin{equation}
C_i(\mu_W)=U(\mu_W,\Lambda)_{ij}C_j(\Lambda),
\label{eq:wc-evolution}
\end{equation}
where $U(\mu_W,\Lambda)$ is the evolutor between the scales $\Lambda$ and $\mu_W$ computed neglecting the effect of SMEFT coefficients in the running of the SM parameters, as consistent with the SMEFT linear approximation, see Ref.~\cite{deBlas:2025xhe} for details. 
The observables used in the fit are computed in terms of the $C_i(\mu_W)$ coefficients at fit time. We do not differentiate between different scales in the few-hundred GeV regime, and use of a common $\mu_W$ also for high-$p_T$ observables, since these observables still have large experimental uncertainties. Since the two-loop SMEFT anomalous dimension is still missing, for consistency we employ tree-level matrix elements at the scale $\mu_W$. 

As far as the SM parameters are concerned, we follow the same approach as in SM EW precision physics and choose to express all observables in terms of a set of input parameters, namely: the Fermi constant $G_F$, the QED fine structure constant constant $\alpha$ or alternatively the W-boson mass $M_W$, the Z-boson mass $M_Z$, the Higgs-boson mass $M_h$, and the lepton masses. In this way, we implicitly trade the values of the SM parameters at the EW scale, including the corrections from the SMEFT, for the chosen set of input values. We notice that, in the quark sector, since the relations between quark masses, CKM angles, and Yukawa couplings are modified by SMEFT contributions, one cannot trade the experimental values of quark masses and CKM angles for the corresponding SM parameters at the EW scale, and keeping the SMEFT contributions in the Yukawa matrices RGE is essential. In the quark flavour sector we therefore treat the Yukawa couplings at the scale $\Lambda$ as fit parameters that are then 
evolved to the EW scale including the effect of the SMEFT in the running. Once evolved to the EW scale, the Yukawa couplings are used to compute the quark masses and CKM angles, that are then fitted to experimental data. 

The SMEFT coefficients in the mass-eigenstate basis are then used to compute the observables at the scale $\mu_W$. To calculate flavour observables, the SMEFT Lagrangian is further matched to the Low Energy Effective Theory (LEFT) using the results of Ref.~\cite{Jenkins:2017jig}, where the massive gauge-bosons, the Higgs-boson and the top-quark fields are integrated out. The LEFT coefficients are then evolved to the energy scale relevant for specific flavour observables using the LEFT RG equations.

For this study we have used the open-source \texttt{HEPfit} framework~\cite{DeBlas:2019ehy,hepfitsite}, where we have
implemented a global fit of the SMEFT including EW, Drell-Yan, Higgs-boson, top-quark, and flavour observables in the \texttt{NPSMEFTd6General} model\footnote{The \texttt{NPSMEFTd6General} model file is available from the \texttt{HEPfit} GitHub repository, which can be accessed through the \texttt{HEPfit} website~\cite{hepfitsite}.}. We use a Bayesian analysis to extract the posterior probability density for the SMEFT coefficients $C_i(\Lambda)$ given the experimental data, a uniform prior for $C_i(\Lambda)$  in the $[-4\pi,4\pi]$ perturbative range which we approximate requiring $C_i(\Lambda)\in[-15,15]$ (while maintaining all the others fixed at zero), and a Gaussian prior for the input parameters $G_F$, $M_W$, $M_Z$, $M_h$, $\alpha_s(M_Z)$ and the lepton masses. SM input parameters and hadronic parameters entering flavour observables are always floated in the fit (see Ref.~\cite{deBlas:2025xhe} for details), while we vary one SMEFT coefficient at a time in the fits presented in this study.
In all fits, \texttt{HEPfit} uses the BAT library~\cite{Caldwell:2008fw} to sample
the probability density function (pdf) of the SMEFT coefficients $C_i(\Lambda)$ via its numerical representation through a Markov Chain Monte Carlo (MCMC). From the MCMC samples, the $68\%$ and $95\%$ Highest Posterior Density Intervals (HPDI) for the SMEFT coefficients are computed. A lower bound on effective NP interaction scale $\Lambda/\sqrt{|C|}$ can then be obtained from the largest absolute value of the $95\%$ HPDI.

The RG evolution from the scale $\Lambda$ to $\mu_W$ is performed within the \texttt{HEPfit} framework using the \texttt{RGEsolver} library~\cite{DiNoi:2022ejg}, while the running in the LEFT is directly implemented in \texttt{HEPfit}. 

The set of Higgs-boson observables considered consists of Higgs-boson signal strengths at 8~TeV from ATLAS and CMS~\cite{ATLAS:2016neq}, as well as the most recent combined measurements of Higgs-boson
production cross sections, decay branching ratios, and simplified template cross sections (STXS) as provided by the ATLAS and CMS collaborations using 139 fb$^{-1}$ of collision data at
$\sqrt{s}=13$~TeV~\cite{ATLAS:2021vrm,ATLAS:2022vkf,ATLAS:2024lyh,ATLAS:2024fkg,CMS:2022dwd,CMS:2025jwz}. These include measurements of Higgs-boson production via gluon-gluon fusion (ggF), vector-boson fusion (VBF), and
associated production with gauge bosons ($ZH$ and $WH$) and top quarks ($t\bar{t}H$ and $tH$) as well as measurements of Higgs-boson decays into $\gamma\gamma, ZZ^*, WW^*,\tau^+\tau^-,b\bar{b},\mu^+\mu^-$, and $Z\gamma$.  
For the STXS we adopt the most recent Stage 1.2 binning and compare with the experimental results presented in~\cite{ATLAS:2024lyh,CMS:2025jwz}. Full correlation matrices as provided by the experimental collaborations are included in the fit. Both cross sections and decay rates are computed at linear order in
the SMEFT coefficients using \texttt{MG5\_aMC@NLO} \cite{Alwall:2014hca} with an in-house UFO file. We do not consider theoretical uncertainties for Higgs observables since they are not at the moment the dominant source of error, although this will have to be reconsidered in future studies. For more details on the non Higgs-boson observables that have been considered in the fit, including how theoretical uncertainties have been considered in each case, we refer to Ref.~\cite{deBlas:2025xhe}.

\section{Results}
\label{sec:results}

\begin{table}[ht]
\centering
\begin{tabular}{|l|c|c|c|c|}
    \hline
  \multicolumn{5}{|c|}{3 TeV}  \\ \hline 
 & \multicolumn{2}{c|}{Full Fit} & \multicolumn{2}{c|}{noH} \\ \hline
& $C_i$ & $\Lambda/\sqrt{|C_i|}$ & $C_i$ & $\Lambda/\sqrt{|C_i|}$ \\ 
\hline
$C_{\phi  \Box}$& $ -0.3 \pm 2.7 $ & $ 1.3 $  & $ -2.1 \pm 6.3 $ & $ 0.8 $  \\ 
$C_{\phi G}$& $ -0.0110 \pm 0.0080 $ & $ 18.3 $  &  $ -2.6 \pm 1.7 $ & $ 1.4 $    \\ 
$C_{\phi B}$& $ -0.007 \pm 0.014 $ & $ 15.8 $  &  $ -3.2 \pm 6.0 $ & $ 0.8 $    \\ 
$C_{\phi W}$& $ -0.025 \pm 0.046 $ & $ 8.8 $  &  $ -2.5 \pm 6.0 $ & $ 0.8 $   \\ 
$C_{\phi WB}$& $ 0.016 \pm 0.026 $ & $ 11.4 $  &  $ 0.027 \pm 0.070 $ & $ 7.3 $    \\ 
$C_{G}$& $ 3.5 \pm 1.8 $ & $ 1.1 $  &  $ 3.1 \pm 2.3 $ & $ 1.1 $    \\ 
$C_{W}$& $ -0.29 \pm 0.44 $ & $ 2.8 $  &  $ -3.0 \pm 5.1 $ & $ 0.9 $    \\ 
\hline \hline
 \multicolumn{5}{|c|}{1 TeV - noRGE }  \\ \hline
& \multicolumn{2}{c|}{Full Fit} & \multicolumn{2}{c|}{noH} \\ \hline
 & $C_i$ & $\Lambda/\sqrt{|C_i|}$ & $C_i$ & $\Lambda/\sqrt{|C_i|}$ \\ 
\hline
$C_{\phi \Box}$                         & $ -0.02 \pm 0.24 $ & $ 1.4 $  &  -- & --  \\
$C_{\phi G}$                            & $ -0.0016 \pm 0.0012 $ & $ 15.8 $  &  -- & --   \\
$C_{\phi B}$                            & $ -0.0007 \pm 0.0014 $ & $ 17.1 $  &  -- & --   \\
$C_{\phi W}$                            & $ -0.0028 \pm 0.0051 $ & $ 8.8 $  &  -- & --   \\
$C_{\phi WB}$                           & $ 0.0014 \pm 0.0026 $ & $ 12.5 $  &  $ 0.0030 \pm 0.0067 $ & $ 7.9 $     \\
$C_{G}$                         & $ 0.07 \pm 0.15 $ & $ 1.6 $  &  $ 0.05 \pm 0.15 $ & $ 1.7 $     \\
$C_{W}$                         & $ -0.50 \pm 0.70 $ & $ 0.7 $  & $ -0.34 \pm 0.77 $ & $ 0.7 $   \\
\hline
\end{tabular}
\caption[]{Results of the individual fits for the $U(3)^5$ flavour symmetric SMEFT, for the operators mainly constrained by Higgs-boson observables. For each coefficient, we report the $68\%$ HPDI interval, the bound on $\Lambda/\sqrt{\vert C_i\vert }$ in units of TeV obtained taking the maximum of the $95\%$ HPDI interval for $\vert C_i \vert$, for the full fit (left) and omitting Higgs-boson observables (right). Results are shown with RG evolution for $\Lambda = 3$ TeV (upper part of the table) and without RG evolution (lower part). A dash denotes that no bound can be set from that particular fit.
\label{tab:u3-ind}  }
\end{table}

In order to illustrate the impact of Higgs-boson observables on full fits of the SMEFT, 
we consider fits varying one SMEFT coefficient at a time, which are best suited to investigate the impact of specific sets of observables. Individual fits also lend themselves better to a first comparison with other groups' results, since each group has used slightly different assumptions that would most certainly affect the complexity of a global fit while have a smaller impact on individual ones. We have organized the results of the fits into two tables, Tables \ref{tab:u3-ind} and \ref{tab:u2-ind}, for the case of $U(3)^5$ and $U(2)^5$ flavour-symmetric NP theories respectively. In the $U(2)^5$ case, we should specify the flavour orientation to define the third generation; however, for the operators considered here, which are mainly constrained by Higgs-boson data, constraints do not depend significantly on the choice of flavour basis (see Table 10 of Ref.~\cite{deBlas:2025xhe}) and we report here results obtained in the basis in which up-type quark Yukawa couplings are diagonal at the scale $\Lambda$ (the so-called "UP basis"). For the purpose of illustrating the impact of RG evolution, we show results obtained for $\Lambda = 3$ TeV including RG evolution (upper part of each table), and results obtained without RG evolution, fixing $\Lambda = 1$ TeV (lower part of each table). Finally, 
to show the impact of Higgs measurements, in each sector we compare the results of the full fit with those obtained by removing Higgs-boson observables.

\renewcommand{\arraystretch}{1.05}
\begin{table}[ht!]
\centering
\footnotesize
\begin{tabular}{|l|c|c|c|c|}
    \hline
  \multicolumn{5}{|c|}{3 TeV}  \\ \hline 
 & \multicolumn{2}{c|}{Full Fit} & \multicolumn{2}{c|}{noH} \\ \hline
& $C_i$ & $\Lambda/\sqrt{|C_i|}$ & $C_i$ & $\Lambda/\sqrt{|C_i|}$ \\ 
\hline
$C_{\phi  \Box}$& $ -0.3 \pm 2.7 $ & $ 1.3 $  & $ -2.1 \pm 6.3 $ & $ 0.8 $  \\ 
$C_{\phi G}$& $ -0.0110 \pm 0.0080 $ & $ 18.3 $  &  $ -2.6 \pm 1.7 $ & $ 1.4 $    \\ 
$C_{\phi B}$& $ -0.007 \pm 0.014 $ & $ 15.8 $  &  $ -3.2 \pm 6.0 $ & $ 0.8 $    \\ 
$C_{\phi W}$& $ -0.025 \pm 0.046 $ & $ 8.8 $  &  $ -2.5 \pm 6.0 $ & $ 0.8 $   \\ 
$C_{\phi WB}$& $ 0.016 \pm 0.026 $ & $ 11.4 $  &  $ 0.027 \pm 0.070 $ & $ 7.3 $    \\ 
$C_{G}$& $ 3.5 \pm 1.8 $ & $ 1.1 $  &  $ 3.1 \pm 2.3 $ & $ 1.1 $    \\ 
$C_{W}$& $ -0.29 \pm 0.44 $ & $ 2.8 $  &  $ -3.0 \pm 5.1 $ & $ 0.9 $    \\ 
$C_{\phi ud}^{[33]}$& $ 5.1 \pm 5.4 $ & $ 0.8 $  & $ 1.5 \pm 8.5 $ & --   \\ 
$C_{e\phi }^{[33]}$& $ 0.017 \pm 0.085 $ & $ 6.8 $  & -- & --  \\ 
$C_{u\phi }^{[33]}$& $ 5.3 \pm 4.3 $ & $ 0.8 $  & -- & --  \\ 
$C_{d\phi }^{[33]}$& $ 0.082 \pm 0.097 $ & $ 6.1 $  & -- & -- \\ 
$C_{eB}^{[33]}$                 & $ \textcolor{red}{6.8 \pm 7.9} $ & \textcolor{red}{--}  & -- & -- \\ 
$C_{uB}^{[33]}$& $ -0.10 \pm 0.17 $ & $ 4.6 $  & $ -0.42 \pm 0.74 $ & $ 2.2 $   \\ 
$C_{eW}^{[33]}$& $ 0.5 \pm 2.3 $ & $ 1.4 $  & -- & --  \\ 
$C_{uW}^{[33]}$& $ -0.19 \pm 0.28 $ & $ 3.5 $  & $ -0.37 \pm 0.71 $ & $ 2.2 $   \\ 
$C_{uG}^{[33]}$& $ -0.27 \pm 0.15 $ & $ 4.0 $  & $ -1.63 \pm 0.67 $ & $ 1.7 $   \\ 
$C_{lequ}^{(1)[3333]}$          & $ -0.12 \pm 0.42 $ & $ 3.1 $& -- & -- \\ 
$C_{lequ}^{(3)[3333]}$          & $ 0.8 \pm 3.0 $ & $ 1.2 $  &  -- & -- \\
\hline \hline
 \multicolumn{5}{|c|}{1 TeV - noRGE }  \\ \hline
& \multicolumn{2}{c|}{Full Fit} & \multicolumn{2}{c|}{noH} \\ \hline
 & $C_i$ & $\Lambda/\sqrt{|C_i|}$ & $C_i$ & $\Lambda/\sqrt{|C_i|}$ \\ 
\hline
$C_{\phi \Box}$                         & $ -0.02 \pm 0.24 $ & $ 1.4 $  &  -- & --  \\
$C_{\phi G}$                            & $ -0.0016 \pm 0.0012 $ & $ 15.8 $  &  -- & --   \\
$C_{\phi B}$                            & $ -0.0007 \pm 0.0014 $ & $ 17.1 $  &  -- & --   \\
$C_{\phi W}$                            & $ -0.0028 \pm 0.0051 $ & $ 8.8 $  &  -- & --   \\
$C_{\phi WB}$                           & $ 0.0014 \pm 0.0026 $ & $ 12.5 $  &  $ 0.0030 \pm 0.0067 $ & $ 7.9 $     \\
$C_{G}(*)$                         & $ 0.07 \pm 0.15 $ & $ 1.6 $  &  $ 0.05 \pm 0.15 $ & $ 1.7 $     \\
$C_{W}$                         & $ -0.50 \pm 0.70 $ & $ 0.7 $  & $ -0.34 \pm 0.77 $ & $ 0.7 $   \\
$C_{\phi ud}^{[33]}(*)$            & $ \textcolor{red}{-8.8 \pm 6.3} $  & \textcolor{red}{--} & $ \textcolor{red}{-8.8 \pm 6.3} $  & \textcolor{red}{--}\\ 
$C_{e\phi }^{[33]}$             & $ 0.0018 \pm 0.0085 $ & $ 7.3 $ & -- & --\\ 
$C_{u\phi }^{[33]}$             & $ 0.51 \pm 0.40 $ & $ 0.9 $ & -- & --\\ 
$C_{d\phi }^{[33]}$             & $ 0.009 \pm 0.012 $ & $ 5.4 $ &  -- & -- \\
$C_{eB}^{[33]}$                 & $ 1.9 \pm 3.2 $ & $ 0.4 $   &  -- & -- \\ 
$C_{uB}^{[33]}$                 & $ 0.024 \pm 0.047 $ & $ 2.9 $  & $ -1.29 \pm 0.94 $ & $ 0.6 $ \\ 
$C_{eW}^{[33]}$                 & $ -3.3 \pm 5.8 $ & \textcolor{red}{--} & -- & --\\ 
$C_{uW}^{[33]}$                 & $ -0.010 \pm 0.075 $ & $ 2.5 $  & $ -0.13 \pm 0.12 $ & $ 1.6 $ \\ 
$C_{uG}^{[33]}$                 & $ 0.009 \pm 0.033 $ & $ 3.7 $   & $ -0.171 \pm 0.083 $ & $ 1.7 $ \\ 
$C_{lequ}^{(1)[3333]}$          & -- & --  &  -- & --\\ 
$C_{lequ}^{(3)[3333]}$          &  -- & --  & -- & --\\ 
\hline
\end{tabular}
\caption[]{Same as Table \protect\ref{tab:u3-ind}, but for the $U(2)^5$ flavour symmetric SMEFT. The coefficients $C_G$ and $C_{\phi ud}^{[33]}$ are only constrained by Higgs-boson observables if RG evolution is included, while without RG effects they can only be constrained by top-quark observables, so that in the noRGE case, which we mark with an asterisk, the fit without Higgs-boson data gives the same result as the full fit. The cases in which the $68\%$ or $95\%$ HPDI intervals touch the edges of the prior distributions are indicated in red.
\label{tab:u2-ind}  }
\end{table}

Starting from the $U(3)^5$ case shown in Table \ref{tab:u3-ind}, we notice that in this flavour scenario, only bosonic operators receive the dominant constraints from Higgs-boson observables. The only purely bosonic operator not constrained by Higgs-boson data is $\mathcal{O}_{\phi D}=\left(\phi^\dag D^\mu \phi\right)^\star \left(\phi^\dag D_\mu \phi\right)$, since it is mainly constrained by EW precision observables~\cite{deBlas:2025xhe}. $C_G$ is also only mildly constrained by Higgs-boson observables, being more sensitive to top-quark observables and in particular constrained by jet observables that however have not been included in our fit. In the context of our fit the most relevant impact of Higgs-boson data is on $C_{\phi G}$, $C_{\phi B}$, and $C_{\phi W}$, which are significantly more constrained in the full fit compared to the fit without Higgs-boson data. Notice that, in the fit without Higgs data, these three operators are only constrained via RG mixing, which is reflected in the absence of bounds in the noRGE case. The other bosonic operators, $C_{\phi \Box}$, $C_{\phi WB}$, $C_G$ and $C_W$, are also more constrained in the full fit, but the impact of Higgs data is less pronounced.

In the $U(2)^5$ case, the individual bounds on bosonic operators are equal to the $U(3)^5$ case, but Higgs-boson data constrain also new operators containing third-generation fermions, in particular $C_{u\phi}^{[33]}$, $C_{d\phi}^{[33]}$ and $C_{e\phi}^{[33]}$, which modify the relation between Yukawa couplings and masses for top, bottom, and tau respectively. In fact, these operators can only be constrained by Higgs-boson observables, since their effect in all other observables can be reabsorbed in a redefinition of quark masses. Furthermore, Higgs-boson data also constrain the dipole operators $C_{uG}^{[33]}$, $C_{uB}^{[33]}$ and $C_{uW}^{[33]}$, which enter in Higgs-boson production via gluon fusion and in the $H\to \gamma\gamma$ and $H\to Z\gamma$ decay rates. The leptonic dipole operators $C_{eB}^{[33]}$ and $C_{eW}^{[33]}$ are also constrained by Higgs-boson data, but the bounds are much weaker. Finally, the four-fermion operator coefficients $C_{lequ}^{(1)[3333]}$ and $C_{lequ}^{(3)[3333]}$ are constrained by Higgs-boson data only via their mixing with dipole operators, so that no bound can be set in the noRGE case.

 Regarding the impact of each channel on a given operator, we observe that, although $t\bar{t}H$ can in principle constrain $C_G$, $C_{u\phi}^{[33]}$ and $C_{uG}^{[33]}$, these operators also contribute to the gluon-fusion production mode which, being much more precisely measured, ultimately provides the strongest bounds. Indeed, gluon fusion also yields the leading constraint on $C_{\phi G}$. The channel that constrains the largest number of operators is $H\to\gamma\gamma$, which provides the leading bounds on $C_{\phi B}$, $C_{\phi W}$, $C_{\phi WB}$, $C_W$, $C_{eB}^{[33]}$, $C_{uB}^{[33]}$, $C_{uW}^{[33]}$ and $C_{eW}^{[33]}$. However, due to the RGE, $C_{eW}^{[33]}$ also receives relevant constraints from $H\to\tau^+\tau^-$, a channel that is crucial to constrain $C_{u\phi}^{[33]}$, $C_{uW}^{[33]}$, $C_{lequ}^{(1)[3333]}$, and $C_{lequ}^{(3)[3333]}$. Finally, another relevant channel is $H\to b\bar{b}$, which provides the leading constraint on $C_{d\phi}^{[33]}$ and $C_{\phi ud}^{[33]}$. A future improvement in this channel would be particularly important to constrain $C_{\phi ud}^{[33]}$, given that the present bound lies around the perturbativity limit. The other channels not explicitly discussed here do not provide the most stringent constraint on any of the operators, although they would become relevant in the context of a global fit, where flat directions can appear. 

\begin{figure}[htb]
    \centering
    \includegraphics[width=0.9\linewidth]{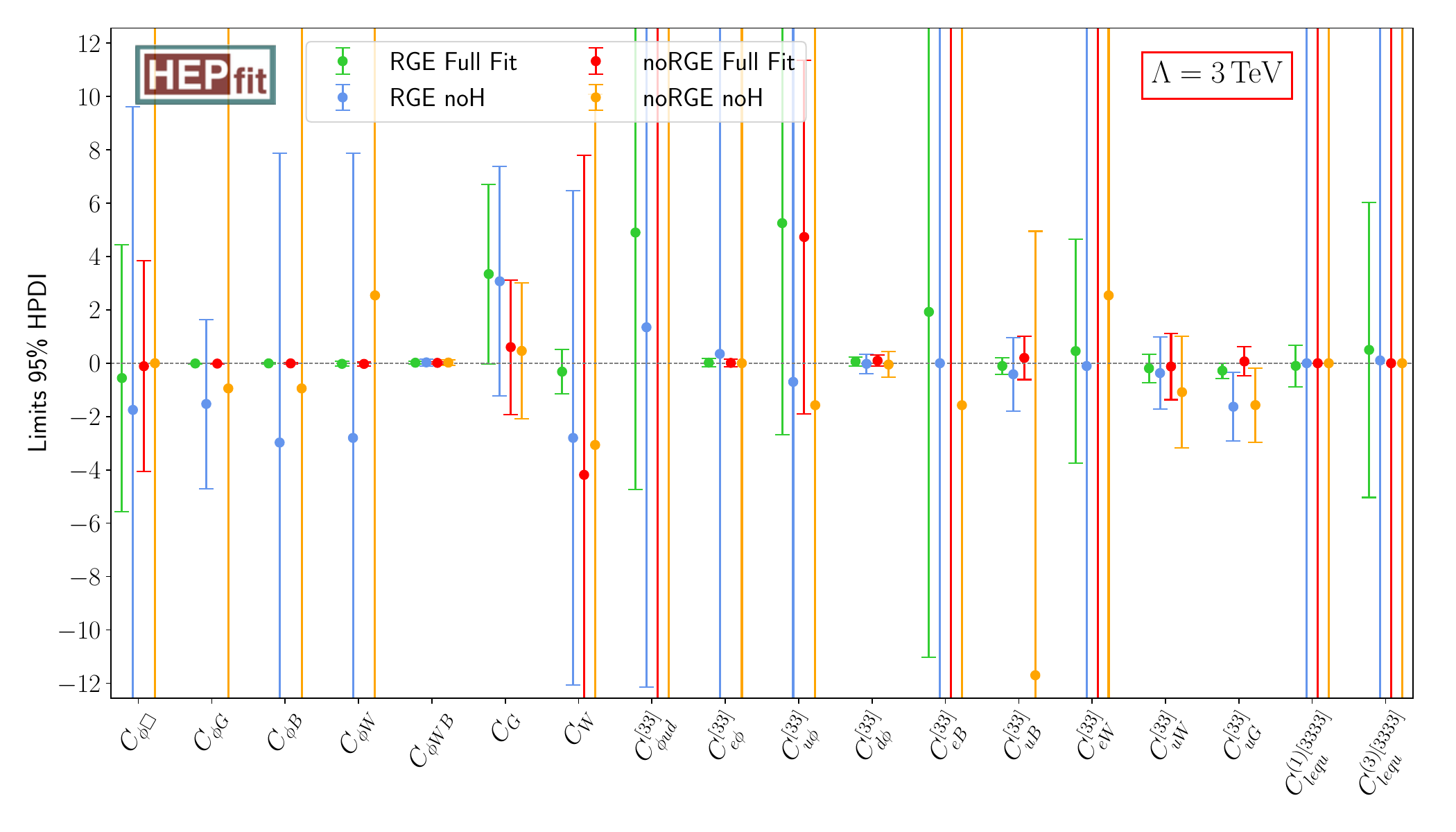}
    \caption{Comparison of constraints from a full fit and a fit that does not include Higgs-boson observables, for the   $U(2)^5$ case. The bounds on the bosonic-operator Wilson coefficients also apply to the $U(3)^5$ case.  The scale of NP has been set to $\Lambda=3$ TeV. The limits shown correspond to the 95\% HPDI. Results are presented with and without the RGE effects (the latter also adjusted to a value of $\Lambda=3$ TeV for this comparison), following the colour scheme indicated in the legend.  }
    \label{fig:u2-dimless}
\end{figure}

\begin{figure}[htb]
    \centering
    \includegraphics[width=0.9\linewidth]{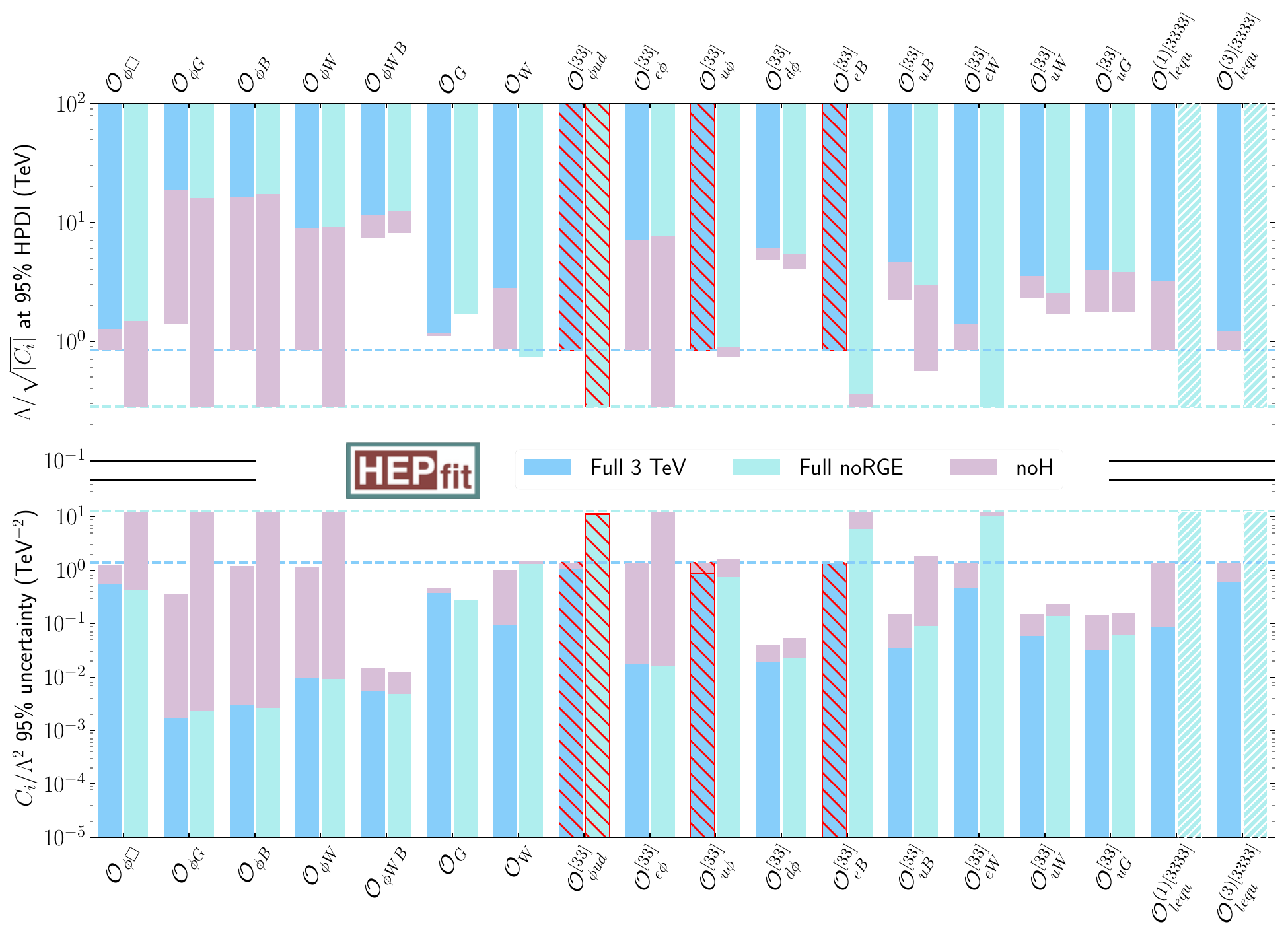}
    \caption{Results from individual fits in the $U(2)^5$ flavour symmetric SMEFT. For each coefficient $C_i$, the top panel shows the scale of NP allowed by the data at 95\% probability (normalized by the square root of the maximum of the 95\% HPDI interval for $|C_i|$). The bottom panel shows the width of the 95\% probability range divided by two. Both panels show results for 1) the full fit with RGE (for $\Lambda=3$~TeV) and without RGE (for $\Lambda=1$~TeV), and 2) the fit that does not include Higgs-boson observables. The color code is as explained in the legend. The horizontal lines indicate the maximum value allowed for each Wilson coefficient in the fit, corresponding to the perturbativity limit $4\pi$ (see Section~\ref{sec:framework}). The cases in which the $95\%$ HPDI interval touches the prior's edges, indicated in red in Table~\ref{tab:u2-ind}, are hatched with red diagonal lines. When the posterior distribution of a coefficient is completely flat the $95\%$ HPDI interval is hatched with diagonal white lines.
    }
    \label{fig:scale-bounds}
\end{figure}

The results of Tables~\ref{tab:u3-ind} and \ref{tab:u2-ind} are illustrated and complemented in Figs.~\ref{fig:u2-dimless} and \ref{fig:scale-bounds} where we report both the bounds on the Wilson coefficients and the corresponding lower bounds on the scale of NP. Including Higgs-boson observables can set lower bounds on $\Lambda$ as high as 15-20~TeV unless correlations exist in given NP models to compensate for the effect of individual operators as illustrated in the SMEFT framework.


\renewcommand{\arraystretch}{1.07}
\begin{table}[ht]
\centering
\begin{tabular}{|l|c|c|c|}
    \hline
  \multicolumn{4}{|c|}{noRGE }  \\ \hline
& HEPfit [This work]
& SMEFiT \cite{Celada:2024mcf} & Ref.~\cite{Bartocci:2023nvp} \\ \hline
$C_{\phi \Box}$                         & $ [-0.49, 0.44]$  & $[-0.4,1.2]$ & $[-1.0,-0.2]$ \\
$C_{\phi G}$                            & $  [-0.0040, 0.0008]$ & $[-0.001,0.005]$  & $[-0.004,0.000]$ \\
$C_{\phi B}$                            & $ [-0.0034, 0.0020] $ & $ [-0.005,0.002]$  & $[-0.003,0.002]$ \\
$C_{\phi W}$                            & $ [-0.013, 0.007] $ & $ [-0.018,0.006] $  & $[-0.009,0.007]$ \\
$C_{\phi WB}$                           & $ [-0.0034, 0.0064] $ & $ [-0.007,0.003] $  & $[-0.004,0.002]$  \\
$C_{W}$                         & $[-1.8, 1.0]$ & $ [-0.61,0.48] $ & $[-0.17,0.36] $ \\
$C_{e\phi }^{[33]}$             & $[-0.015, 0.019]$ & $[-0.027,0.036] $ &\\ 
$C_{u\phi }^{[33]}$             & $[-0.3, 1.3]$ & $ [-1.2,0.3] $&\\ 
$C_{d\phi }^{[33]}$             & $ [-0.013, 0.035] $ & $ [-0.007,0.04] $ &\\
$C_{uB}^{[33]}$                 & $ [-0.07, 0.12] $ & $ [-0.07,0.22] $  &\\ 
$C_{uW}^{[33]}$                 & $ [-0.16, 0.14] $ & $ [-0.029,0.087] $  & \\ 
$C_{uG}^{[33]}$                 & $ [-0.055, 0.073] $ & $ [-0.10,-0.005] $  & \\ 
\hline
\end{tabular}
\caption[]{Comparison of the results of the individual fits for selected operators with those of Refs.~\cite{Celada:2024mcf} and \cite{Bartocci:2023nvp}. For each coefficient, we report the $95\%$ HPDI (or CL for Ref.~\cite{Bartocci:2023nvp}) interval for $C_i/\Lambda^2$ (in TeV$^{-2}$). Results are obtained without RG evolution. For Ref.~\cite{Bartocci:2023nvp} we only report the non-vanishing coefficients under the $U(3)^5$ flavour assumption adopted in that study. 
\label{tab:SMEFiT}  }
\end{table}
Let us now compare our results with the literature. 
Ref.~\cite{Celada:2024mcf} presents results for a slightly different flavour assumption with respect to the ones considered in this study. A $U(2)_q\times U(2)_u \times U(3)_d$ symmetry is assumed in the quark sector, where, in addition to the top quark, also the bottom and charm Yukawa couplings are kept nonzero to account for the current LHC sensitivity to these parameters. $\left(U(1)_l\times U(1)_e \right)^3$ is assumed in the lepton sector, also in this case keeping the tau-lepton Yukawa coupling non-zero as motivated by LHC data. Although not exactly the same, this flavour assumption is quite close to the $U(2)^5$ case considered in this study when focusing on operators mainly constrained by Higgs-boson measurements. It should be noted, however, that Ref.~\cite{Celada:2024mcf} does not include the full set of operators analysed in this work.  While RG evolution is present in their following work \cite{terHoeve:2025gey}, only global fit results are present for that case so in order to compare the results of the individual fits we use Ref.~\cite{Celada:2024mcf}, which does not include RG evolution.\footnote{We have nevertheless compared with Ref.~\cite{terHoeve:2025gey} where RGE/noRGE results are presented in Table 3.4 and we find overall good qualitative agreement.} 
\\
Ref.~\cite{Bartocci:2023nvp} presents results for a $U(3)^5$ flavour symmetry, as in our first scenario, but does not include RG evolution. 
While RG effects are included in the subsequent analysis presented in Ref.~\cite{Bartocci:2024fmm} only constraints from the global fit obtained by profiling the other coefficients are reported there, so for the comparison of individual fits, as done for the previous collaboration, we stick to Ref.~\cite{Bartocci:2023nvp} . 

In Table \ref{tab:SMEFiT} we compare the results of our individual fits with those of Refs.~\cite{Celada:2024mcf} and \cite{Bartocci:2023nvp} for the operators reported in those analyses which are mainly constrained by Higgs-boson observables when neglecting RG evolution. Considering the different data sets used and the different fitting procedures, the results are in good agreement. Tracking the origin of the small differences observed would require a detailed comparison of the different implementations, which is beyond the scope of this study.

\section{Conclusion}
\label{sec:conclusion}
We have presented the state of the art on the impact of Higgs-boson measurements on the determination of SMEFT coefficients, showing that Higgs-boson observables are giving the dominant constraint on several operators, providing a sensitivity on the scale of NP well above $10$ TeV, driven mainly by $C_{\phi G}$, $C_{\phi W}$, $C_{\phi B}$ and $C_{\phi WB}$. In the future, the HL-LHC will significantly improve the precision of Higgs-boson measurements~\cite{ATL-PHYS-PUB-2025-018} and hopefully improve the sensitivity of all the operators listed in Table \ref{tab:smeft-operators} and/or show hints of NP. On the theoretical side, the ongoing effort to improve to next-to-leading order the SMEFT anomalous dimension, together with the computation of one-loop amplitudes for the observables of interest, will allow to strongly reduce the theoretical uncertainty in the calculation of Higgs-boson observables and provide quite strong constraints on NP theories as well as possible indirect indications of physics beyond the SM. The analysis presented in this study, which considers different UV scales as starting points of the RGE, shows that the LO RGE is already fairly stable. Nevertheless, the NLO RGE should not only provide a more accurate account of Wilson coefficients' scale dependence, but also introduce new effects from operators that first enter either in the 2-loop anomalous-dimension mixing or in the one-loop matrix elements. Moving forward, important steps to be considered by the Higgs/EFT working group may include: a more detailed comparison of existing fit results leading to a validation of the different fitting frameworks; a systematic study of the impact of improved experimental and/or theoretical precision, as well as of additional observables; a discussion of possible extensions of the considered set of coefficients, for example including CP violating effects (see Ref.~\cite{Barducci:2025ati} for a recent LHC Higgs WG study) or additional operators beyond the dimension-six SMEFT framework. 



\paragraph{Funding information}
The work of J.B. has been partially funded by 
the FEDER/Junta de Andaluc\'ia project grant P18-FRJ-3735,  MICIU/AEI/\allowbreak10.13039/501100011033 and FEDER/UE (grant PID2022-139466NB-C21). The work of V.M. is supported by the MICIU through a Beatriz Galindo Junior grant (BG24/00038), the European Research Council (ERC) under the
European Union’s Horizon 2020 research and innovation programme (Grant agreement No. 949451) and a Royal Society University Research Fellowship through grant URF/R1/201553.
The work of L.R. is supported in part by the U.S. Department of Energy under grant DE- SC0010102 and by
INFN through a Foreign Visiting Scientist Fellowship. This work was supported in part by the European Union - Next Generation
EU under Italian MUR grant PRIN-2022-RXEZCJ. L.S. and M.V. acknowledge support from the project ``Exploring New Physics'' funded by INFN.







\bibliography{hepfit.bib}


\end{document}